\documentclass[twocolumn,prl]{revtex4}
\usepackage{graphics,graphicx}
\begin{document}
\title{Chemically induced ferromagnetism near room temperature in single crystal (Zn$_{1-x}$Cr$_{x}$)Te half-metal}
\author{J. Guo$^{1}$}
\author{A. Sarikhani$^{2,\dagger}$}
\author{P. Ghosh$^{1}$}
\author{T. Heitmann$^{3}$}
\author{Y. S. Hor$^{2}$}
\author{D. K.~Singh$^{1,*}$}
\affiliation{$^{1}$Department of Physics and Astronomy, University of Missouri, Columbia, MO}
\affiliation{$^{2}$Department of Physics, Missouri University of Science and Technology, Rolla, MO}
\affiliation{$^{3}$University of Missouri Research Reactor, Columbia, MO}
\affiliation{$^{*}$email: singhdk@missouri.edu}

\begin{abstract}

Magnetic semiconductors are at the core of recent spintronics research endeavors. Chemically doped II-VI diluted magnetic semiconductors, such as (Zn$_{1-x}$Cr$_{x}$)Te, provide promising platform in this quest. However, a detailed knowledge of the microscopic nature of magnetic ground state is necessary for any practical application. Here, we report on the synergistic study of (Zn$_{1-x}$Cr$_{x}$)Te single crystals using elastic neutron scattering measurements and density functional calculations. For the first time, our research unveils the intrinsic properties of ferromagnetic state in macroscopic specimen of (Zn$_{0.8}$Cr$_{0.2}$)Te. The ferromagnetism is onset at $T_c \sim$ 290 K and remains somewhat independent to modest change in the substitution coefficient x. We show that magnetic moments on Zn/Cr site develop ferromagnetic correlation in a-c plane with large ordered moment of $\mu$ = 3.08 $\mu_B$. Magnetic moment across the lattice is induced via the mediation of Te site, uncoupled to the number of dopant carriers as inferred from the density functional calculation. Additionally, the ab-initio calculations also reveal half-metallicity in x = 0.2 composition. These properties are highly desirable for future spintronic applications.  
 
\end{abstract}

\maketitle
Diluted magnetic semiconductors (DMS) are attractive for spintronic applications, as the underlying ferromagnetic and electrical properties can be easily tuned by modest chemical doping.\cite{Furdyna,Mcdonald,Kurodai} Previous efforts in this regard have mainly focused on the exploration of III-V DMS compounds, such as Mn-doped GaAs and InAs.\cite{Nazmul,Munekata,Ohno,Akai,Konig} In these materials, the ferromagnetic transition temperature, Curie temperature, is sufficiently far below room temperature, which is one of the key requisites for the spintronic application. So far, the highest detected Curie temperature in these compounds is $T_c \sim$ 170 K in Mn $\delta$-doped GaAs/Be-doped p-type AlGaAs heterostructures.\cite{Nazmul} Also, both magnetic and electrical phenomena are interlinked and directly attributed to the carrier (hole) doping of Mn ions-- holes ferromagnetically mediate between d-orbital local moments.\cite{Konig,Van,Mcdonald} Therefore, higher doping of Mn ions, needed to enhance the ferromagnetic Curie temperature in III-V DMS, would inevitably change the semiconducting property into metallic characteristic. 

In contrast to the III-V DMS, exploration of II-VI diluted magnetic semiconductors e.g. (Zn$_{1-x}$Cr$_{x}$)T where T = Se, S, Te have revealed the persistence of near room temperature ferromagnetism.\cite{Saito,Pekarek,Patel} Additionally, theoretical and experimental study of II-VI DMS have demonstrated the uncoupled nature of magnetic and electrical phenomena.\cite{Saito,Mac,Larson,Rhyne} In this case, magnetism is derived from the exchange interaction between the delocalized s, p band electrons and localized d-electrons of magnetic ions, not directly from the magnetic impurity e.g. Cr.\cite{Mac} A positive value of exchange interaction would indicate ferromagnetic state of the system.\cite{Furdyna} Experimental investigation of s, p-d exchange interaction using magnetic circular dichroism measurements have revealed positive exchange constant in (Zn$_{1-x}$Cr$_{x}$)Te thin film.\cite{Saito}

Among the many compositions of (Zn$_{1-x}$Cr$_{x}$)T, tellurides are of special importance. ZnTe is a wide band gap nonmagnetic semiconductor with energy gap of $\Delta \sim$ 6 eV.\cite{Chadi} It crystallizes in the zinc-blend cubic structure with lattice parameter of a = 6.103 $\AA$.\cite{Mac} The compound manifests near room temperature ferromagnetism at modest Cr doping in thin film specimen of (Zn$_{1-x}$Cr$_{x}$)Te (x$\sim$ 0.035) despite a very low carrier density of 1.10$^{15}$ cm$^{-3}$.\cite{Saito2}. More recently, study of polycrystalline (Zn$_{1-x}$Cr$_{x}$)Te has confirmed the occurrence of high temperature ferromagnetism.\cite{Ali} It was also inferred that the DMS material at 20\% Cr substitution exhibits half-metallic characteristic, which can be used to generate the fully spin-polarized current.\cite{Ali} Interestingly, similar conclusions were drawn in the recent density functional calculations of analogous DMS material (ZnCr)Se.\cite{Zhang} Although, these properties are of strong technological importance, but the lack of fundamental understanding of the ground state magnetic configuration in (Zn$_{1-x}$Cr$_{x}$)Te hinders practical application. We have synthesized high quality single crystal samples of various substitution coefficients that are used to elucidate the intrinsic magnetic properties. Details about the quality of single crystals can be found elsewhere.\cite{Ali}

\begin{figure}
\centering
\includegraphics[width=8.6 cm]{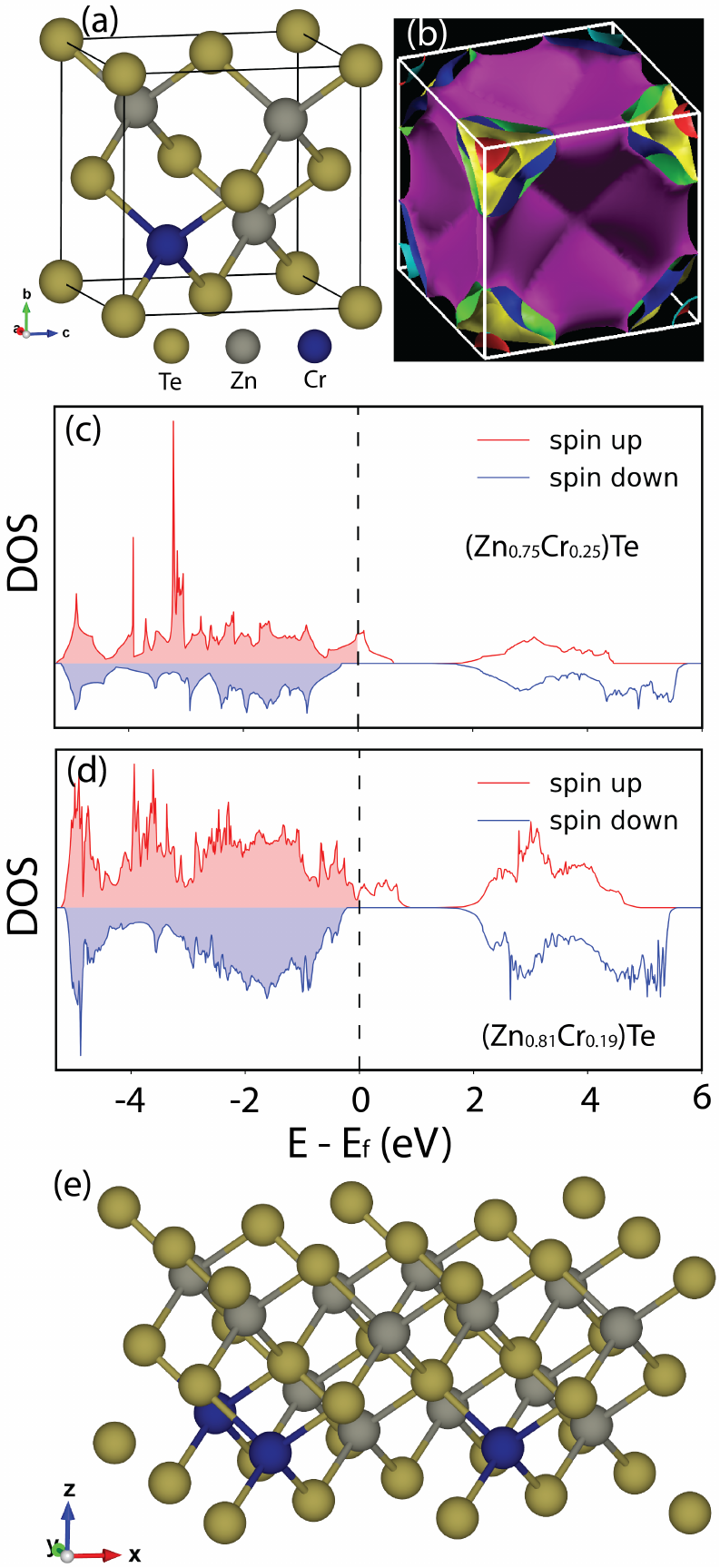} \vspace{-6mm}
\caption{(color online) Density functional theory calculations. (a) Chemical structure of (Zn$_{0.75}$Cr$_{0.25}$)Te, used for the DFT calculations. (b) Density functional theory calculated Fermi surface, viewed from the top and tilted-axes projection. Parallel surfaces in the hole pockets are separated by wave vector, $\overline{{\bm{\alpha\gamma}}} =$ (100) rlu in the reduced units of 2${\pi}/a$ for the cubic structure, also consistent with experimental finding. (c) Spin polarized density of states in 25\% Cr-substituted ZnTe, depicting half-metallic state. (d) Similar results are obtained in DFT calculations on (Zn$_{0.81}$Cr$_{0.19}$)Te. (e) Illustration of (Zn$_{0.81}$Cr$_{0.19}$)Te used in the DFT calculation.
} \vspace{-6mm}
\end{figure}

In this article, we report on the synergistic investigation of two substitution coefficients, x = 0.2, 0.15, using the ab initio calculation and detailed elastic neutron scattering measurements of single crystal specimens. We find direct evidence of a long-range ferromagnetic order with transition temperature of $T_c$ = 290 K in the bulk material. The Curie temperature does not change for a modest variation in the Cr substitution percentage. Surprisingly, the power-law exponent of the order parameter, $\beta$ = 0.46(02), in x = 0.2 composition is larger than that typically found in ferromagnetic materials. Such a large $\beta$ value suggests inclination to the critical behavior of correlated moments. Numerical modeling of the experimental data illustrates ferromagnetic alignment of moments in the a-c plane with an ordered moment of $\mu \sim$ 3.08 $\mu_B$. The estimated ferromagnetic moment is also consistent with the ab-initio density functional theory (DFT) calculations. Given the fact that most research works have focused on the magnetic study of epitaxial thin film of (ZnCr)Te where substrate effect and reduced dimensionality of the specimen can be playing important roles,\cite{Soundararajan,Kuroda} this is the first time we unequivocally elucidate the intrinsic magnetic nature of the chemically doped (Zn$_{1-x}$Cr$_{x}$)Te. 

\begin{figure}
\centering
\includegraphics[width=8.6 cm]{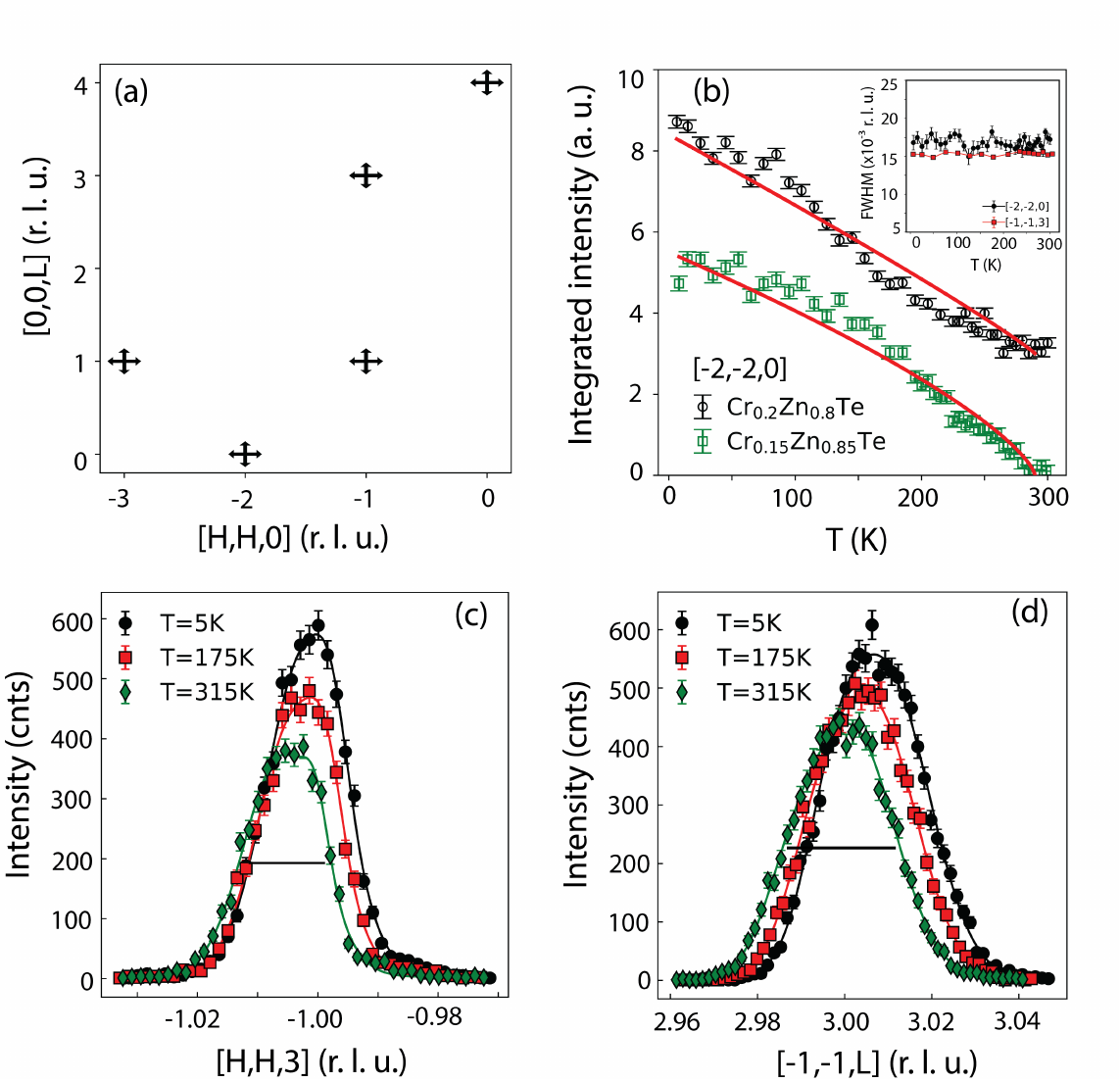} \vspace{-6mm}
\caption{(color online) Neutron scattering measurement of (Zn$_{1-x}$Cr$_{x}$)Te, x = 0.2, 0.15. (a) Schematic of elastic scans across the nuclear Bragg peak positions in multiple Brillouin zones. Measurements are performed in both HH and L directions to accurately determine the integrated intensity. (b) Magnetic order parameters of (Zn$_{1-x}$Cr$_{x}$)Te, x = 0.2, 0.15 as a function of temperature. Magnetic order parameter is obtained by subtracting the high temperature data, at $T$ = 325 K, to consecutive low temperature data. The intensity was estimated by fitting individual rocking scan at (-1-13) peak position at different temperatures. Fitting of order parameter by power law (see text for detail) yields $T_c$ = 290 K in both x = 0.2 and 0.15 substitution coefficients with the power law exponent of 0.46(02) and 0.36(04), respectively. Inset shows the temperature dependence of FWHM. (c-d) Representative scans of neutron measurements on (Zn$_{0.8}$Cr$_{0.2}$)Te at different temperatures along HH- and L- directions across (-1-13) rlu. As temperature reduces, nuclear peak intensity increases; thus, suggesting the development of ferromagnetic ground state. Experimental data are well described by Gaussian lineshape. In all plots, error bars represent one standard deviation.
} \vspace{-4mm}
\end{figure}

The density functional theory calculations are carried out using a plane-wave basis set as implemented in the QUANTUM-ESPRESSO software. \cite{ Giannozzi}  The simulated system, illustrated in Fig. 1a, is constructed by replacing one Zn atom with Cr atom per unit cell, corresponding to 25\% doping level. The Generalized Gradient Approximation (GGA) along with the revised Perdew-Burke-Ernzerhof exchange-correlation functional (PBEsol) are implemented with projector-augmented wave (PAW) pseudopotentials.\cite{John,Kresse} A well-converged kinetic energy cutoff of 80 Ry for wave function and charge density of 640 Ry are used throughout the calculation. Geometry optimization is performed by relaxing the atomic positions via systematically changing the lattice parameter with energy and force convergence thresholds of 0.0001 and 0.001 a. u., respectively. Brillouin zone (BZ) integrations during the structure optimization are performed on a Monkhorst-Pack k-grid of 8 $\times$ 8 $\times$ 4, while a uniform dense grid of 30 $\times$ 30 $\times$ 30 is used for the construction of the Fermi surface (FS) and the density of state (DOS).\cite{ Monkhorst,Timrov} The DFT+U+V approach is adopted to account for the localization effect in the Cr 3d manifold as well as the hybridization between Cr 3d manifold and Te 5p manifold. The on-site interaction U of Cr 3d manifold and the inter-site interaction V between Cr 3d manifold and Te 5p manifold are determined by using a q-grid of 5 $\times$ 5 $\times$ 5 within the density functional perturbation theory (DFPT), as implemented in the QUANTUM-ESPRESSO software.\cite{Timrov} The DFT calculation yields ferromagnetism in this system. The calculated moment on Cr and Te sites are $\sim$ 4 $\mu_B$ and 0.1 $\mu_B$, respectively. The ordered moment values are also consistent with other reports. Moments are antiparallel with each other on neighboring Cr and Te sites.\cite{Wang,Shoren} The calculated spin-resolved density of states (DOS), presented in Fig. 1c, show that the Fermi energy passes through the spin-up DOS while there is a gap of the spin-down DOS, confirming the half-metallic nature of this system.\cite{Wang,Shoren} DFT calculation of $\sim$ 19\% doping level yields similar results, see Fig. 1d. The simulation was carried out on a 2 $\times$ 2 $\times$ 1 zinc-blende 32-atom ZnTe supercell, where three Zn atoms are replaced by Cr atoms such that nearest neighbor Cr atom pairs are avoided. Three hole pockets are observed in the spin-up Fermi surface, shown in Fig. 1b, which is consistent with the p-type doping of Cr in the host ZnTe structure. Parallel surfaces are separated by (100) rlu along the crystallographic directions.

\begin{figure}
\centering
\includegraphics[width=8.7 cm]{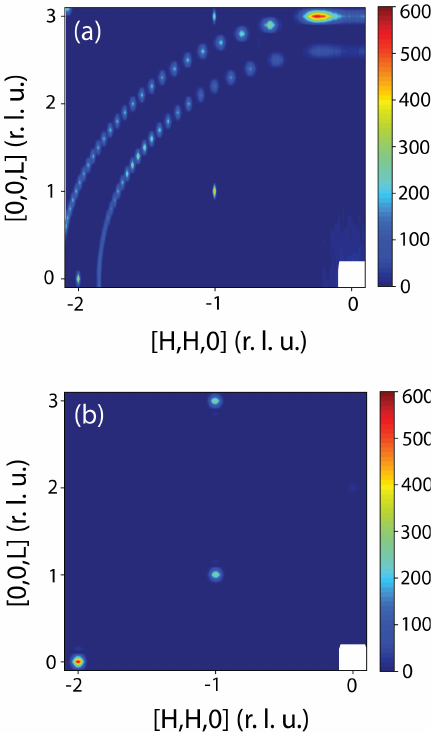} \vspace{-6mm}
\caption{(color online) Numerical modeling of experimental data. (a) Detailed color map, depicting nuclear and magnetic scattering across several Brillouin zones, obtained on TRAIX spectrometer at $T$ = 5 K. The circular streaks are arising due to aluminum powder lines from the sample holder. (b) Numerical modeling of experimental data was independently performed using model calculations. Calculated scattering pattern for moment configuration shown in Fig. 4 describes experimental data.
} \vspace{-6mm}
\end{figure}

Experimental verification of the intrinsic magnetism in (Zn$_{0.8}$Cr$_{0.2}$)Te is obtained from elastic neutron scattering measurements. Neutron scattering measurements were performed on a 8 mg flux grown thin rectangular shape single crystal, of dimension 3 mm (length) $\times$ 2mm (width) $\times$ 1.25 mm (thickness), at the thermal Triple Axis Spectrometer, TRIAX, at the University of Missouri Research Reactor (MURR). Elastic measurements were performed at the fixed final energy of 14.7 meV using PG (pyrolitic graphite) monochromator. The measurements on TRIAX employed a flat pyrolytic graphite (PG) analyzer with collimator sequence of PG filter-60'-60'-Sample-40'-PG filter-40'. Single crystal sample was mounted at the end of the cold finger of a closed cycle refrigerator with a base temperature of $T \simeq$ 5 K. Measurements were performed with the crystal oriented in the ($HHL$) scattering plane. Here, $H$ and $L$ represent reciprocal lattice units of 2${\pi}/a$ and 2${\pi}/c$, respectively, with c = a. 

Single crystal allows for a detailed examination of the intensities and magnetic scattering pattern, which reveals the nature of spin correlations that are not possible to obtain from magnetic and thermodynamic measurements. Elastic scans were obtained along both $HH$- and $L$-crystallographic directions, as shown schematically in Fig. 2a. In Fig. 2c-d, we show representative scans at various temperatures across [-1-13] nuclear Bragg peak along $HH$ and $L$- directions. As the sample is cooled to low temperature, scattering intensity enhancement of nuclear Bragg peak becomes apparent. Neutron scattering data is well described by the resolution convoluted Gaussian line shape. Full width at half maximum at representative peaks are shown in the inset of Fig. 2d. We see that the width of the peak does not change as a function of temperature. Also, it is comparable to the resolution of the thermal triple axis spectrometer. The additional scattering, which is arising due to magnetic correlation in the sample, indicates the development of commensurate long range magnetic order in the system. At low temperature, all nuclear peaks are found to manifest enhanced elastic intensity, limited by the form factor of Cr atom. 

\begin{figure*}
\centering
\includegraphics[width=17cm]{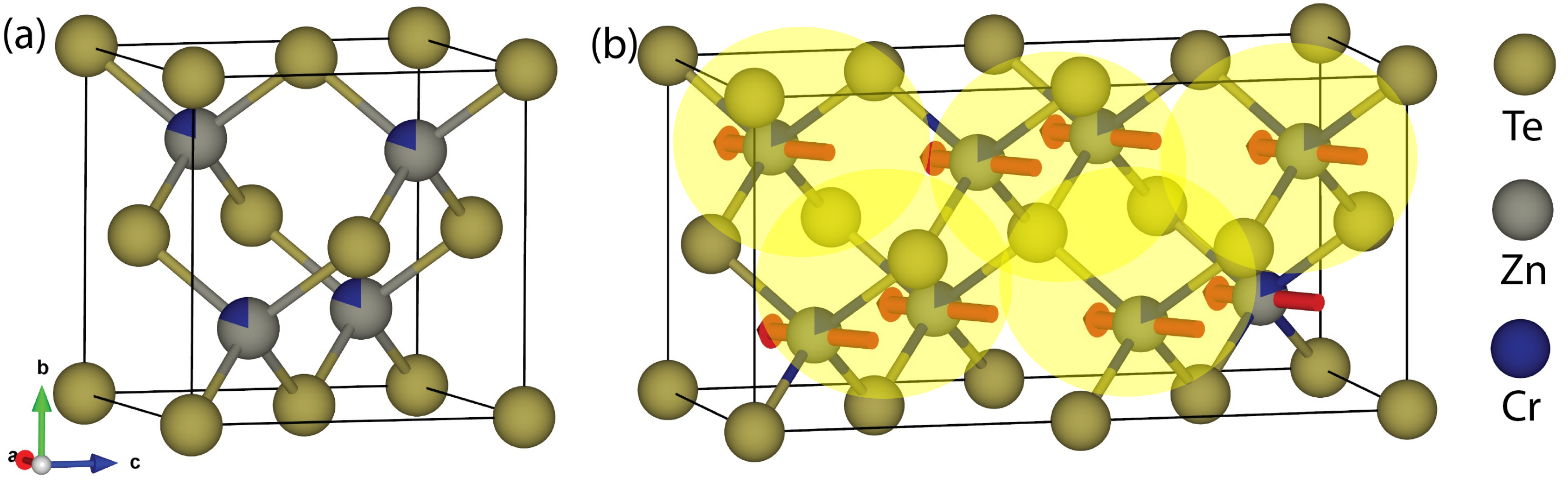} \vspace{-2mm}
\caption{(color online) Ground state spin correlation and induced magnetism. Refinement using FullProf-SARAh representational analysis suggests ferromagnetic arrangement of magnetic moments on Zn/Cr site with polar angles of $\theta$ = 120$^{o}$ and $\phi$ = -2$^{o}$. The large estimated moment, arguably, arises due to the induced magnetism mediated by orbital overlap on Te-sites, depicted by transparent circles.  
} \vspace{-6mm}
\end{figure*}

The chemical substitution of Zn by Cr induces magnetism via the modification of the Fermi surface in (Zn$_{1-x}$Cr$_{x}$)Te, as evidenced by the DFT calculations. In fact, recent studies have revealed that the magnetism in (ZnCr)Te is onset at a doping percentage of as low as 5\% of Cr atoms.\cite{Saito2,Ali} Besides the detailed experimental investigation of 20\% Cr substitution level, we have also performed neutron measurements on (Zn$_{0.85}$Cr$_{0.15}$)Te compound. Magnetic order parameter as a function of temperature for both 15\% and 20\% Cr substitutions are plotted in Fig. 2b. The order parameter data are fitted with the power law equation: $ I \propto  (1-\frac{T}{T_{c}})^{2\beta} $, to accurately estimate the magnetic transition temperature and the critical exponents that can provide important information about the nature of the phase transition. Fitting of experimental data using the power law equation yields Curie temperature of $T_c \sim$ 290 K in both compounds. The value of exponent beta for the substitution coefficients of x = 0.2 and 0.15 are estimated to be 0.46(02) and 0.36(04), respectively, that suggest the three-dimensional interaction in the ordered regime. These beta values are very close to the critical phase transition exponent $\sim$ 0.5. We also observe unusually linear-type trend in the order parameter plot of both compounds, albeit more pronounced in higher substitution coefficient of x = 0.2. Typically, a ferromagnetic transition is accompanied by the first order phase transition. The magnetic phase transition in (ZnCr)Te clearly departs from the conventional trend. Perhaps, the nature of magnetic moment correlation can shed light on this.

To understand the nature of ground state spin correlation in (Zn0.8Cr0.2)Te, detailed maps of nuclear and magnetic scattering peaks across several Brillouin zones were obtained at two temperatures of $T$ = 325 K and $T$ = 5 K on TRIAX spectrometer. We show the color plot of mesh scan data at $T$ = 5 K in Fig. 3a. Stronger magnetic scattering is observed along the [00L] direction, compared to the [HH0] positions. Since magnetic contribution to the total intensity at structural Bragg peaks are more pronounced at low temperature, we subtract the high temperature data to the low temperature data for numerical modeling purposes. The numerical modeling of experimental results are performed using the FullProf program in conjunction with the representational analysis by SARAh program.\cite{SARAh} In this case, the number of symmetry-allowed magnetic structures, possible for a particular crystallographic site, is simply the number of non-zero irreducible representations in the magnetic representation.The best fit to experimental data is found for a ferromagnetic configuration of magnetic moments on Zn/Cr sites that align along the diagonal direction in the $a-c$ basal plane, defined by polar coordinates of $\theta$ = 120$^{o}$ and $\phi$ = -2$^{o}$, see Fig. 4. The estimated size of ordered moment is $\mu \sim$ 3.08 $\mu_B$.

The ground state spin configuration was independently verified using the model calculations. The simulated scattering intensities are obtained by adding up both the nuclear scattering
and the magnetic scattering contribution. For the simulation purposes, we have used multiple lattice units of the size 10$\times$10$\times$10. Also, a weighted average scattering length is used for the Zn/Cr sites. The magnetic structure factor is calculated using the formula of $ F_{M} =\sum_{j} S_{\perp j} p_{j}  e^{ iQr_{j} }e^{-W_{j}} $,\cite{Shirane} where $ S_{\perp} =\hat{Q}\times(S\times\hat{Q})$ is the spin component perpendicular to the Q, $ p = (\frac{\gamma r_{0}}{2})gf(Q) $, $ (\frac{\gamma r_{0}}{2}) $= 0.2695 $ \times 10^{-12} cm$, g is the Lande splitting factor and was taken to be g = 2, $ f(Q) $ is the magnetic form factor and $e^{-W_{j}} $ is the Debye-Waller factor and was taken to be 1. 20\% of Zn sites are randomly substituted by Cr atoms in the lattice units for the calculation of the magnetic structure factor. The simulated pattern is shown in Fig 3b. Numerically simulated scattering pattern for the spin configuration, shown in Fig. 4, well describes experimental results for the ordered moment of $\mu \sim$ 3.08 $\mu_B$.

Such a large ordered moment can be arising due to the clustering of Cr ions or the induced magnetism on Zn/Cr site, mediated by Te ions. The clustering of Cr ions would result in the finite size effect, often reflected by the short-range order in elastic measurements, as found in spin glass systems.\cite{Singh} But this is not what we observe. Resolution limited elastic peaks suggest the presence of long range magnetic order in the system. So, magnetism in Cr-doped compound is most likely associated to the induced magnetism mediated by the orbital overlap between Zn/Cr and Te sites. Subsequently, we expect other compositions with varying Cr percentages to manifest same or similar magnetic structure, albeit with smaller (larger) ordered moment size at small (large) Cr substitution. The argument of induced magnetism is also consistent with the DFT calculations from a string of reports, including ours as discussed above. Given the fact that 20\% substitution of Zn by Cr would result in the random replacement of approximately one Zn atom by the Cr atom per unit cell, the size of ordered moment cannot be this large. Cr ion in ZnTe has 3$d$$^{4}$ electronic configuration. Therefore, even in the case of entire unit cell being occupied by Cr ion (instead of Zn/Cr ion), the maximum ordered moment may not be larger than 4 $\mu_B$. Hence, (Zn$_{0.8}$Cr$_{0.2}$)Te is expected to exhibit a much smaller ordered moment than this optimum value as less than one site per unit cell, on the average, is occupied by the Cr atom. The only explanation to this conundrum lies in the phenomena of induced magnetism. A similar behavior arises in (GaMn)As where Mn substitution induces large magnetic moment across the entire unit cell, even though the parent compound GaAs is non-magnetic.\cite{Mcdonald}

In summary, we have performed detailed synergistic investigation of underlying magnetism in single crystal specimen of Cr-doped ZnTe using DFT calculation and neutron scattering measurements. Previous study of (ZnCr)Te have mainly focused on the epitaxially grown thin film samples of reduced dimensionality where the substrate can have pronounced effect on physical and magnetic properties. The study of bulk crystalline material unveils the intrinsic property of the system. Our research works reveal the ferromagnetic nature of moment correlation along the diagonal direction in a-c plane. The long-range ferromagnetic order is onset at high temperature of $T_c$ = 290 K. Also, the Curie temperature does not seem to vary for a modest change in the Cr substitution percentage. Importantly, the ordered moment on Zn/Cr site is quite large, $\mu$ = 3.08 $\mu_B$, despite the small substitution coefficient in (Zn$_{0.8}$Cr$_{0.2}$)Te. We argue that such large, ordered moment arises due to the induced magnetism between Zn/Cr-Te orbital overlaps and not depends on the number of dopant carriers. The qualitative explanation as well as the size of ordered moment are in good agreement with the DFT calculations. Further research works, elucidating the quantitative aspect of exchange interaction in (ZnCr)Te using inelastic neutron scattering measurements, are highly desirable. It will require large single crystal samples that are currently not available. DFT calculations also suggest half-metallicity in (Zn$_{0.8}$Cr$_{0.2}$)Te. The theoretical finding, although consistent with recent experimental report,\cite{Ali} needs further investigation via electrical transport measurements. The chemically induced ferromagnetic property in (ZnCr)Te semiconductor can have practical implication to the spintronics research. Finding a suitable semiconducting material with chemically tunable ferromagnetism, which persists to room temperature or higher, is an important research problem. (ZnCr)Te can provide a strong platform in this endeavor.  

\section{}

\textbf{Acknowledgements}

DKS thankfully acknowledges the support by the Department of Energy, Office of Science, Office of Basic Energy Sciences under the grant no. DE-SC0014461.

\textbf{Data availability statement}

The data that support the findings of this study are available from the corresponding author upon request.

$^{\dagger}$Author contributed equally

\clearpage

\end{document}